\documentstyle[12pt]{article}
\begin{document}
\topmargin -0.2cm \oddsidemargin -0.2cm \evensidemargin -1cm
\textheight 22cm \textwidth 12cm

\title{New Fundamental \emph{Light Particle} and Breakdown of Stefan-Boltzmann's Law}
\author{V.N. Minasyan  and V.N. Samoylov \\
Scientific Center of Applied Research, JINR,\\
 Dubna, 141980, Russia}

\date{\today}

\maketitle

\begin{abstract}
Recently, we predicted the existence of fundamental particles in Nature, neutral
\emph{Light Particles} with spin 1 and rest mass $m=1.8\times 10^{-4} m_e$,
in addition to electrons, neutrons and protons. We call these
particles Light Bosons because they create electromagnetic field
which represents Planck's gas of massless photons together with a
gas of \emph{Light Particles} in the condensate. Such reasoning leads to a breakdown of
Stefan--Boltzmann's law at low temperature. On the other hand,
the existence of new fundamental neutral \emph{Light Particles} leads to correction of such physical concepts
as Bose–Einstein condensation of photons, polaritons and exciton polaritons.
\end{abstract}

\vspace{100mm}

\vspace{5mm}

\section{Introduction}

First, the quantization scheme for the local electromagnetic field in
vacuum was presented by Planck in his black body radiation
studies [1]. In this context, the classical Maxwell equations lead
to appearance of the so-called ultraviolet catastrophe; to remove
this problem, Planck proposed the model of the electromagnetic field as
an ideal Bose gas of massless photons with spin one. However,
Dirac [2] showed the Planck photon-gas could be obtained through a
quantization scheme for the local electromagnetic field, presenting
a theoretical description of the quantization of the local
electromagnetic field in vacuum by use of a model Bose-gas of local
plane electromagnetic waves propagating by speed $c$ in vacuum.

In a different way, in regard to Plank and Dirac's models, we consider the
structure of the electromagnetic field [3] as a non-ideal gas consisting
of $N$ neutral \emph{Light Bose Particles} with spin 1 and
finite mass $m$, confined in a box of volume $V$. The form of
potential interaction between \emph{Light Particles} is defined by
introduction of the principle of wave-particle duality of de
Broglie [4] and principle of gauge invariance. In this respect,
a non-ideal Bose-gas consisting of \emph{Light Particles} with spin $1$
and non-zero rest mass is described by Planck's gas
of massless photons together with a gas consisting of \emph{Light Particles}
in the condensate. In this context, we defined the
\emph{Light Particle} by the model of hard sphere particles [5].
Such definition of \emph{Light Particles} leads to cutting off the spectrum of
the electromagnetic wave by the boundary wave number $k_0=\frac{m
c}{\hbar}$ or boundary frequency $\omega_{\gamma}=10^{18}$ Hz of gamma
radiation at the value of the rest mass of the \emph{Light Particle}
$m=1.8\times 10^{-4} m_e$. On the other hand, the existence of the boundary
wave number $k_0=\frac{m c}{\hbar}$ for the electromagnetic field in
vacuum is connected with the characteristic length of the
interaction between two neighboring \emph{Light Bosons} in the coordinate
space with the minimal distance $d=\frac{1}{k_0}=\frac{\hbar}{m
c}=2\times 10^{-9}m$. This reasoning determines the density of \emph{Light Bosons}
$\frac{N}{V}$ as $\frac{N}{V}=\frac{3}{4\pi d^3}=0.3\times
10^{26}m^{-3}$.

It is well known that Stefan-Boltzmann's law [6] for
thermal radiation, presented by Planck's formula [1], determines the average
energy density  $\frac{U}{V}$ as
\noindent
\begin{equation}
\frac{U}{V}= \frac{2}{V}\sum_{ 0\leq k< \infty}\hbar k
c\overline{\vec{i}^{+}_{\vec{k}} \vec{i}_{\vec{k}}}=\sigma T^4,
\end{equation}

\noindent
where $\hbar$ is the Planck constant; $\sigma$ is the Stefan–Boltzmann constant;
$\overline{\vec{i}^{+}_{\vec{k}}\vec{i}_{\vec{k}}}$ is the average
number of photons with the wave vector $\vec{k}$ at the temperature
$T$:

\begin{equation}
\overline{\vec{i}^{+}_{\vec{k}}\vec{i}_{\vec{k}}}=
\frac{1}{e^{\frac{\hbar k c}{kT}}-1}
\end{equation}
Obviously, at $T=0$, the average energy density vanishes in Eq.(1),
i.e.$\frac{U}{V}=0$, which follows from Stefan–Boltzmann's law.

However, as we show, the existence of the predicted \emph{Light Particles} breaks
Stefan–Boltzmann's law for black body radiation at low temperature.

\section{Breakdown of Stefan-Boltzmann's law}

Now, we consider the results of letter [3], where the average energy density
of black radiation $\frac{U}{V}$ is represented as:

\begin{equation}
\frac{U}{V}= \frac{mc^2 N_{0, T} }{V}+\frac{2}{V}\sum_{ 0< k<
k_0}\hbar k c\overline{\vec{i}^{+}_{\vec{k}} \vec{i}_{\vec{k}}},
\end{equation}

where $\frac{mc^2 N_{0, T} }{V}$ is a new term, in regard to Plank's
formula (1), which determines the energy density of \emph{Light Particles} in
the condensate; $\frac{N_{0, T} }{V}$ is the density of \emph{Light Particles}
in the condensate.

In this respect, the equation for the density of \emph{Light Particles} in
the condensate $\frac{N_{0, T} }{V}$ represents as:

\begin{equation}
\frac{N_{0, T} }{V}=\frac{N }{V}- \frac{1}{V}\sum_{0< k<
k_0}\frac{L^2_{\vec{k}}}{1-L^2_{\vec{k}}}- \frac{1}{V}\sum_{ 0< k<
k_0}\frac{1+L^2_{\vec{k}}}{1-L^2_{\vec{k}}}
\overline{\vec{i}^{+}_{\vec{k}}\vec{i}_{\vec{k}}}
\end{equation}

with the real symmetrical function $L_{\vec{k}}$ from the wave vector
$\vec{k}$:

\begin{equation}
L^2_{\vec{k}}=\frac{\frac{\hbar^2 k^2}{2m }+ \frac{mc^2}{2}-\hbar k
c}{\frac{\hbar^2 k^2}{2m }+ \frac{mc^2}{2}+\hbar k c}.
\end{equation}

Our calculation shows that at absolute zero the value of
$\overline{\vec{i}^{+}_{\vec{k}} \vec{i}_{\vec{k}}}=0$, and
therefore the average energy density of black radiation
$\frac{U}{V}$ reduces to

\begin{equation}
\frac{U}{V}= \frac{mc^2 N_{0, {T=0}} }{V}=\frac{mc^2 N }{V}-
\frac{m^4 c^5 B(2,3)}{4\pi^2 \hbar^3}\approx \frac{mc^2 N }{V},
\end{equation}

where $B(2,3)=\int^{1}_{0}x(1-x)^2 dx=0.1$ is the beta
function.

Thus, the average energy density of black radiation
$\frac{U}{V}$ is a constant at absolute zero. In fact, there is a
breakdown of Stefan-Boltzmann's law for thermal radiation.

In conclusion, it should be also noted that \emph{Light Bosons} in vacuum create photons, while \emph{Light Bosons} in a homogeneous medium generate the so-called polaritons. This fact implies that photons and polaritons are quasiparticles, therefore, Bose–Einstein condensation of photons [7], polaritons [8] and exciton polaritons [9] has no physical sense.

\section*{Acknowledgements}

We are particularly grateful to Professor Marshall Stoneham F R S
(London Centre for Nanotechnology, and Department of Physics and
Astronomy University College London, UK) as well as Elena Aleksandrovna Nikulnikova (Joint Institute for Nuclear Research, Dubna, Russia) for help with the English.

\end{document}